\documentclass[epj]{webofc}
\usepackage[varg]{txfonts}   
\newcommand{\Det}{{\rm Det}}
\newcommand{\Tr}{{\rm Tr}}
\newcommand{\tr}{{\rm tr}}

\newcommand{\be}{\begin{equation}}
\newcommand{\ee}{\end{equation}}
\newcommand{\ba}{\begin{array}{l}}
\newcommand{\ea}{\end{array}}

\newcommand{\re}[1]{(\ref{#1})}

\woctitle{ICNFP 2017}
\begin{document}
\selectlanguage{english}
\title{Gluons, Heavy and Light Quarks in the QCD Vacuum}

\author{Mirzayusuf Musakhanov\inst{1}\fnsep\thanks{\email{musakhanov@gmail.com}} 
}

\institute{National University of Uzbekistan}

\abstract{
We are discussing the properties of the QCD vacuum which might be
important especially for the understanding of hadrons with small quark
core size $\sim0.3\,fm.$ We assume that at these distances  the
QCD vacuum can be described by the Instanton Liquid Model (ILM). At
larger distances, where confinement is important, ILM should be extended
to Dyons Liquid Model (DLM). The ILM has only two free parameters,
average instanton size $\rho\approx0.3\,fm$ and average inter-instanton
distance $R\approx1\,fm$, and can  successfully describe the key
features of light hadron physics. One of the important conceptual
results was prediction of the momentum dependent dynamical quark mass
$M\sim(packing\,\,fraction)^{1/2}\,\,\rho^{-1}\approx360\,MeV$,
later confirmed numerically by evaluations in the lattice. The estimates
show that gluon-instanton interaction strength is also big and is
controlled by the value of dynamical gluon mass $M_{g}\approx M$.
Heavy quarks interact with instantons much weaker. The heavy quark-instanton
interaction strength is given by $\Delta m_{Q}\sim packing\,\,fraction\,\,\rho^{-1}\approx70\,MeV.$
Nevertheless, the direct instanton contribution to the colorless heavy-heavy
quarks potential is sizable and must be taken into account. At small
distances, where one-gluon exchange contribution to this potential
is dominated, we have to take into account dynamical gluon mass $M_{g}$.
Also, instantons are generating light-heavy quarks interactions and
allow to describe the nonperturbative effects in heavy-light quarks
systems. 
 
}
\maketitle

\section{Introduction}

\label{intro}

QCD instanton is a topologically nontrivial classical solution of
Yang-Mills (YM) equations for gauge fields in Euclidean space, which
is a tunneling path between Chern-Simons (CS) states \cite{BPST1975}.
Then, within quantum mechanics QCD vacuum can be considered as the
lowest  energy quantum state of the one-dimensional crystal along
the collective CS coordinate \cite{FJR1976}.

Without any doubt instantons represent a very important topologically
nontrivial component of the QCD vacuum. In ILM the external classical
gluon field is given by 
\begin{equation}
A_{\mu}=\sum_{I}A_{\mu}^{I}(\gamma_{I}),\label{eq:sum}
\end{equation}
where $A_{\mu}^{I}(\gamma_{I})$ is a generic notation for the QCD
(anti)instanton in the singular gauge, described by its collective
coordinates $\gamma_{I}$ (the position in Euclid 4D space $z_{I}$,
the size $\rho_{I}$ and the $SU(N_{c})$ color orientation $U_{I}$,
$4N_{c}$ variables altogether ). The main parameters of the QCD instanton
vacuum ( Instanton Liquid Model(ILM)) are the average instanton size
$\rho$ and inter-instanton distance $R$ (see reviews \cite{Diakonov:2002fq,Schafer:1996wv}).
Their values were phenomenologically estimated as 
\begin{equation}
\rho=1/3\,fm,\,R=1\,fm\label{eq:SetInstanton}
\end{equation}
 and confirmed by theoretical variational calculations \cite{Diakonov:2002fq,Schafer:1996wv}
and recent lattice simulations of the QCD vacuum \cite{lattice}.
 
\begin{figure}[h]
\centering \includegraphics[scale=0.25]{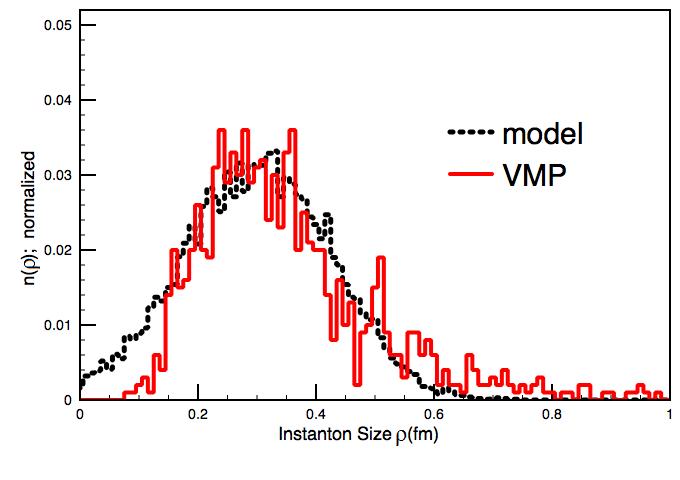} \caption{Instanton size distribution function $n(\rho)$ \textendash{} lattice
vs ILM \cite{instantonsize}.}
\label{instantonsize} 
\end{figure}

The instanton size distribution $n(\rho)$ 
 has been studied independently by lattice simulations (see Fig.\ref{instantonsize}).
As we can see, for large-size instantons the density $n(\rho)$ is
suppressed, which justifies the use of ansatz~(\ref{eq:sum}). However,
the  the large size tail of distribution function $n(\rho)$ becomes
important in the confinement regime of QCD. In this regime (as well
as for temperature $T>0$), instead of individual instanton sum~(\ref{eq:sum})
 we have to replace BRST instantons by KvBLL dyon-instantons \cite{KvBLL1998}
described in terms of  dyons. So, we get an extension of ILM \textendash{}
Liquid Dyon Model (LDM)\cite{Diakonov2009,Shuryak2015}, which is
able to reproduce confinement\textendash deconfinement. Small size
instantons can be still described in terms of their collective coordinates.
The average size of instantons in LDM is $\bar{\rho}\approx0.5\,{\rm fm}$
\cite{Diakonov2009,Shuryak2015}, while in ILM $\bar{\rho}\approx0.3\,{\rm fm}$.

The instanton vacuum background~(\ref{eq:sum}) leads to nonzero
QCD vacuum energy density $\epsilon\approx-500\,MeV/fm^{3}$ \cite{Schafer:1996wv}
and spontaneous breakdown of chiral symmetry~\cite{Goeke:2007bj}
which plays a pivotal and significant role in describing the lightest
hadrons and their interactions.

\begin{table}[h]
\centering \caption{Quarkonium states and its sizes in non-relativistic potential model
\cite{quarkoniumsize}.}
\label{Quarkonium states and its sizes} \scalebox{0.9}{ %
\begin{tabular}{|c|c|c|c||c|c|c|c|c|}
\hline 
State  & $J/\psi$  & $\chi_{c}$  & $\psi'$  & $\Upsilon$  & $\chi_{b}$ & $\Upsilon'$ & $\chi_{b}'$ & $\Upsilon^{''}$ \tabularnewline
\hline 
mass {[}Gev{]}  & 3.07  & 3.53  & 3.68  & 9.46  & 9.99 & 10.02 & 10.26 & 10.36 \tabularnewline
\hline 
size $r$ {[}fm{]}  & 0.25  & 0.36  & 0.45  & 0.14  & 0.22 & 0.28 & 0.34 & 0.39 \tabularnewline
\hline 
\end{tabular}} 
\end{table}

For applications of the ILM to the physics of the heavy quarks, we
may notice that the typical sizes of quarkonia are $r_{J/\psi}=0.47\,fm,\,\,\,r_{\Upsilon}=0.2\,fm$~\cite{Eichten:1979ms}.
Similar estimate of nucleon quark core size gives $r_{N}\sim0.3-0.5$
fm \cite{nucleon-quarkcore}. Since small quark core size hadrons
are insensitive to the confinement we may safely apply  ILM for their
description.

\section{Light quarks in ILM}

\label{Light quarks} Zero modes are the solutions of the Dirac equation
$(\hat{p}+g\hat{A}_{\pm})\Phi_{\pm,0}(x,\zeta_{\pm})=0.$ Their dominance
in the single instanton light quark propagator provide summation of
the multi-scattering series for the ILM light quark propagator and
leads to the low-frequencies part of the light quarks partition function
\cite{Diakonov:2002fq,Schafer:1996wv,Musakhanov,Goeke:2007bj} 
\begin{eqnarray}
Z[\xi^{+},\xi] & = & \int D\zeta\,\Det_{low}(\hat{p}+g\hat{A}+im)\exp{(-\xi^{+}(\hat{p}+g\hat{A}+im)^{-1}\xi)}=\\
 &  & =\int D\zeta\prod_{f}D\psi_{f}D\psi_{f}^{\dagger}\exp\int\left(\psi_{f}^{\dagger}(\hat{p}\,+\,im_{f})\psi_{f}+\psi_{f}^{\dagger}\xi_{f}+\xi_{f}^{+}\psi_{f}\right)\nonumber \\
 &  & \times\prod_{f}\left\{ \prod_{+}^{N_{+}}V_{+,f}[\psi^{\dagger},\psi]\prod_{-}^{N_{-}}V_{-,f}[\psi^{\dagger},\psi]\right\} ,\nonumber \\
V_{\pm,f}[\psi^{\dagger},\psi] & = & i\int dx\left(\psi_{f}^{\dagger}(x)\,\hat{p}\Phi_{\pm,0}(x;\zeta_{\pm})\right)\int dy\left(\Phi_{\pm,0}^{\dagger}(y;\zeta_{\pm})(\hat{p}\,\psi_{f}(y)\right).
\end{eqnarray}
where $\psi^{\dagger},\psi$ correspond to constituent quarks, $\Phi_{\pm,0}$are
the zero modes in the single instanton and antiinstanton fields respectively.

Small packing fraction $(\rho/R)^{4}\approx 0.01$ justifies
independent averaging over collective coordinates $\zeta$ of each
instanton and leads to the non-local t'Hooft-like vertex with $2N_{f}$-quarks
legs  
\begin{eqnarray}
\overline{V_{\pm}[\psi^{\dagger},\psi]}=\int d\zeta_{\pm}\prod_{f}V_{\pm,f}\left[\psi^{\dagger},\psi\right].
\end{eqnarray}

\begin{figure}[h]
\centering \includegraphics[scale=0.35]{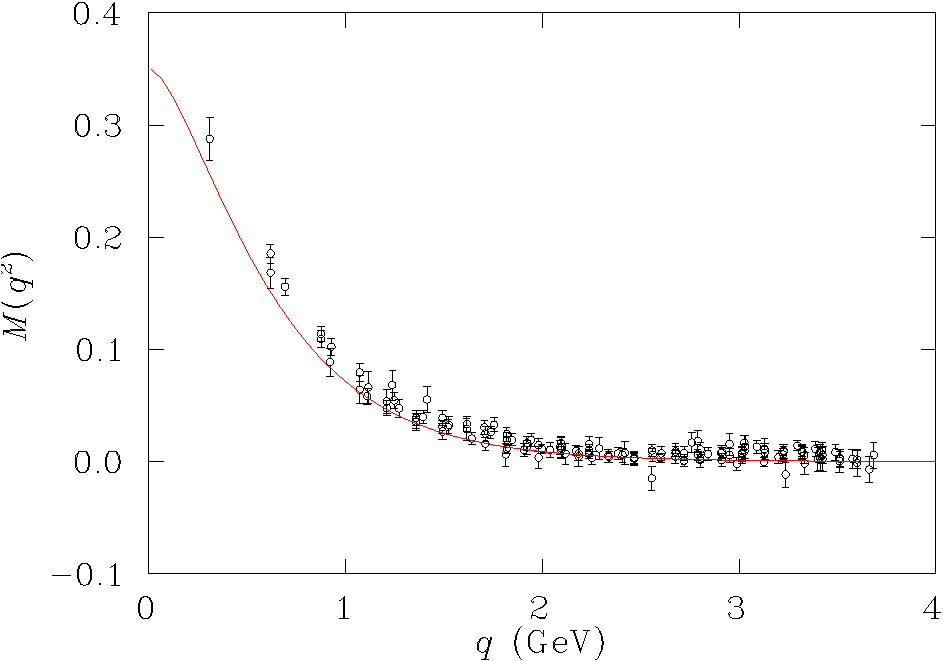} \caption{Dynamical light quark mass M(q). ILM at $\rho=0.33\,fm,R=1\,fm$ vs
lattice \cite{Bowman2005}.}
\label{M(q)} 
\end{figure}

\textbf{Spontaneous Breaking of the Chiral Symmetry.} 
The calculation of $Z[\xi^{+},\xi]$ in the saddle-point approximation
(leading order in $1/N_{c}$) leads to the Spontaneous Breaking of
the Chiral Symmetry (SBCS). One of the manifestations of the SBCS
is the dynamical quark mass $M(q)$ (see Fig.\ref{M(q)}). 

For small momenta $M(0)\approx360$~MeV and is related to the strength
of light quark-instanton interaction. Even in the  leading order over
$1/N_{c}$, the ILM successfully reproduces quark condensate, pion
and nucleon properties etc (see e.g. \cite{Diakonov:1995qy,Diakonov:2002fq,Schafer:1996wv}).
In the next to leading order $1/N_{c}$ corrections , we have successfully
reproduced  the Low Energy Constants of ChPT \cite{Goeke:2007bj}.

\section{Heavy quarks in ILM}

\label{Heavy quarks} It is known that for the description of heavy
quark systems we can develop a systematic $1/m_{Q}$expansion. In
the leading order the heavy quark propagator is given by $w=\int D\zeta(\theta^{-1}-iA_{4})^{-1},$
where $<t_{2}|\theta|t_{1}>=\theta(t_{2}-t_{1})$ and $<T|w|0>=\int D\zeta\,P\exp(i\int_{L}dx_{4}A_{4}),\,\,\,L=(\vec{x},0,T).$
The problem of the averaging over instantons was solved in \cite{DPP1989}
using the framework suggested in~\cite{Pobylitsa1989} , which leads
to $w^{-1}=\theta^{-1}+\sum_{i}\int d\zeta_{i}\left(w-a_{i}^{-1}\right)^{-1}.$

In view of the low density of the instanton liquid, it makes sense
ot develop a systematic expansion over the  dimensionless effective
parameter--the packing fraction $\rho^{4}N/V=\rho^{4}/R^{4}\approx 0.01$.
To the lowest order in the instanton density the heavy quark propagator
is given by 
\begin{eqnarray}
w^{-1}=\theta^{-1}-\frac{N}{2}\tr_{c}\sum_{\pm}\theta^{-1}(w_{\pm}-\theta)\theta^{-1}+O(N^{2}/V^{2}),\label{w}
\end{eqnarray}
where we introduced shorthand notation for the single (anti)instanton
heavy quark propagator $w_{\pm}=\int d\zeta_{\pm}\left(\theta^{-1}-iA_{\pm,4}\right)^{-1}$.
Instanton medium contribution to the heavy quark mass is given by~\cite{DPP1989,Chernyshev:1995gj}
\begin{eqnarray}
\Delta m_{Q}=16\pi i_{0}(0)(\rho^{4}/R^{4})\rho^{-1}/N_{c},\,\,\,i_{0}(0)=0.55,\label{deltamQ}
\end{eqnarray}
where form-factor $i_{0}(x)$ was defined at \cite{DPP1989}. We estimate
that the value of $\Delta m_{Q}$ should be within  $\Delta m_{Q}\approx70\,\mathrm{MeV}$
(for the set of parameters~(\ref{eq:SetInstanton})) and   $\Delta m_{Q}\approx140\,\mathrm{MeV}$
(for $\rho=0.36\,\mathrm{fm},R=0.89\,\mathrm{fm}$). $\Delta m_{Q}$
is related to the strength of a heavy quark-instanton interaction.

\textbf{Colorless state heavy quark-antiquark potential in ILM. } Static
central and spin-dependent parts of the heavy quark-antiquark potential
can be obtained from ILM averaged Wilson loop over rectangular contour
$\vec{r}\times T$ with $T\rightarrow\infty$ \cite{Eichten1980}.
The application of the framework~\cite{DPP1989,Turimov:2016adx}
gives the potential 
\begin{eqnarray}
V(\vec{r})=V_{C}(r)+V_{SS}(r)(\vec{S}_{Q}\!\cdot\!\vec{S}_{\bar{Q}})+V_{LS}(r)(\vec{L}\cdot\vec{S})+V_{T}(r)\left[3(\vec{S}_{Q}\!\cdot\!\vec{n})(\vec{S}_{\bar{Q}}\!\cdot\!\vec{n})-\vec{S}_{Q}\cdot\vec{S}_{\bar{Q}}\right],\label{potential1}
\end{eqnarray}
where $\vec{L}$ is the angular moemntum of the rlative motion, $\vec{S}_{Q,\bar{Q}}$
are the spins of the quarks, and 
\begin{eqnarray}
V_{SS}(r)=\frac{1}{3m_{Q}^{2}}\nabla^{2}V_{C}(r),V_{LS}(r)=\frac{1}{2m_{Q}^{2}}\frac{1}{r}\frac{dV_{C}(r)}{dr},V_{T}(r)=\frac{1}{3m_{Q}^{2}}\left(\frac{1}{r}\frac{dV_{C}(r)}{dr}-\frac{d^{2}V_{C}(r)}{dr^{2}}\right).\label{potential2}
\end{eqnarray}
\begin{figure}[h]
\includegraphics[scale=0.34]{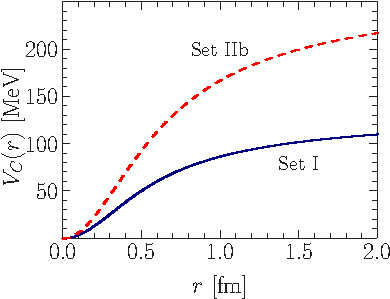} \includegraphics[scale=0.34]{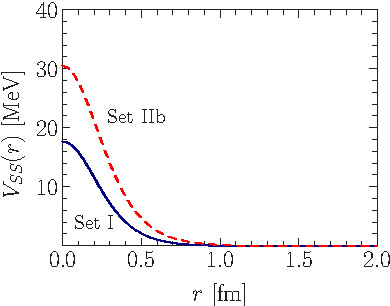}
\includegraphics[scale=0.34]{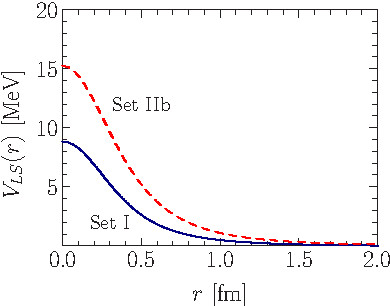} \includegraphics[scale=0.34]{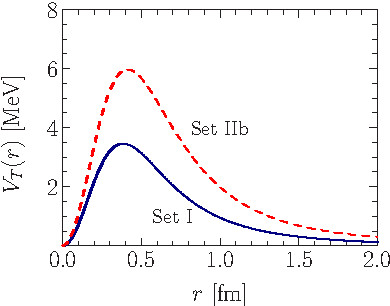}
\caption{Solid curve $\sim$ Set~I $\rho=0.33$~fm and ${R}=1\,\mathrm{fm}$~\cite{Diakonov:2002fq,Schafer:1996wv}.
Dashed one $\sim$ Set~II $\rho=0.36$~fm, ${R}=0.89\,\mathrm{fm}$~\cite{lattice,Goeke:2007bj},
$m_{c}=1275$~MeV.}
\label{potential3} 
\end{figure}

The ILM contributions to the mass shifts of the charmonium states
\cite{Turimov:2016adx} are given in the Table \ref{charmonium}:

\begin{table}[h]
\centering \caption{ILM contribution to the charmonium states. $\Delta M_{c\bar{c}}=M_{c\bar{c}}-2m_{c}$
in {[}MeV{]}.}
\label{charmonium} %
\begin{tabular}{c|c|c|c}
$\Delta M_{c\bar{c}}(J^{P})$ & Set~I & Set~II  & Exp.\tabularnewline
\hline 
$\Delta M_{\eta_{c}}(0^{-})$  & 118,81  & 203,64  & $433,6\pm0.6$\tabularnewline
$\Delta M_{J/\psi}(1^{-})$  & 119,57  & 205,36  & $546,916\pm0.11$\tabularnewline
$\Delta M_{\chi_{c0}}(0^{+})$  & 142,43  & 250,86  & $864,75\pm0.31$\tabularnewline
\end{tabular}
\end{table}

We can see that the instanton effects are not small $\sim30-40\,\%$
in comparison with the experimental data and strongly depend on instanton
liquid parameters.

\section{Gluons in ILM}

\label{gluon} 
\textbf{Scalar \char`\"{}gluons\char`\"{} in ILM.} 
There  operators $\Delta_{I}^{-1}=(p+A_{I})^{2}$ and $\Delta^{-1}=(p+\sum_{i}A_{i})^{2}$
do not have zero mdoes, for this reason their inverse operators are
well-defined,  
\begin{eqnarray}
\Delta=(p^{2}+\sum_{i}(\{p,A_{i}\}+A_{i}^{2})+\sum_{i\neq j}A_{i}A_{j})^{-1},\,\,\,\Delta_{i}=(p^{2}+\{p,A_{i}\}+A_{i}^{2})^{-1},\,\,\,\Delta_{0}=p^{-2}.\label{delta}
\end{eqnarray}
where $\tilde{\Delta}=(p^{2}+\sum_{i}(\{p,A_{i}\}+A_{i}^{2}))^{-1}.$

The propagator in ILM is $\bar{\Delta}\equiv<\bar{\Delta}>=\int D\zeta\,\Delta.$
Let's start first with evaluation of $\bar{\tilde{\Delta}}.$ The
extension of Pobylitsa equation to the present case is 
\begin{eqnarray}
\bar{\tilde{\Delta}}^{-1}-\Delta_{0}^{-1}=\sum_{i}<\{\bar{\tilde{\Delta}}+(\Delta_{i}^{-1}-\Delta_{0}^{-1})^{-1}\}^{-1}>\label{tildedelta}
\end{eqnarray}
Since effective parameter of instanton density expansion in fact is
$\rho^{4}/R^{4}\sim(1/3)^{4}=0.012$, we may neglect  the higher orders
of the expansion and we have at the first order in the instanton density
\begin{eqnarray}
\bar{\tilde{\Delta}}^{-1}-\Delta_{0}^{-1}=N\Delta_{0}^{-1}(\bar{\Delta}_{I}-\Delta_{0})\Delta_{0}^{-1},\label{1ordertildedelta}
\end{eqnarray}
It is obvious that with the same accuracy we have $\bar{\Delta}=\bar{\tilde{\Delta}}$.

\textbf{Scalar \char`\"{}gluon\char`\"{} dynamical mass in ILM.} Applying
the well-known result for the $\Delta_{I}$ \cite{BrownCarlitz},
we found 
\begin{eqnarray}
M_{s}(q)=\left[\frac{3\rho^{2}}{(N_{c}^{2}-1)R^{4}}4\pi^{2}\right]^{1/2}q\rho K_{1}(q\rho),\label{Ms}
\end{eqnarray}
which was also obtained in \cite{hutter}. For the set of parameters~(\ref{eq:SetInstanton})
we got an estimate $M_{s}(0)=256\,MeV$ and the form-factor $q\rho K_{1}(q\rho)$
given by Fig.\ref{ff}: 
\begin{figure}[h]
\centering \includegraphics[scale=0.5]{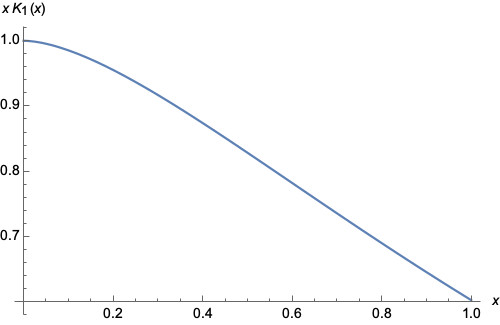} \caption{the form-factor $q\rho K_{1}(q\rho)$}
\label{ff} 
\end{figure}

\textbf{Gluons. Zero-modes problem.} 
The quadratic form of the gluon fluctuations $a_{\mu}$ in the background
of a single instanton is $(a_{\mu}M_{\mu\nu}^{I}a_{\nu})$. The matrix
$M_{\mu\nu}^{I}$ has $4N_{c}$ zero-modes $M_{\mu\nu}^{I}\phi_{\nu}^{i}=0$,
which correspond to the fluctuations along of the collective coordinates
$\zeta_{I}$. If we define $P_{\mu\nu}^{I}$ as the  zero-modes projection
operatorthen, the single instanton gluon propagator $S_{\mu\nu}^{I}$
is given by 
\begin{eqnarray}
M_{\mu\nu}^{I}S_{\nu\rho}^{I}=\delta_{\mu\nu}-P_{\mu\nu}^{I}\label{SI}
\end{eqnarray}
The explicit solution for Eq.\re{SI} was found in~\cite{BrownCarlitz}.

Now, in order to extend Pobylitsa Equation,  we introduce artificial
gluon mass $m$ and define additional quantities $G_{m,\rho\nu}^{I}$
and $g_{m,\mu\nu}^{I}$ defined as~\cite{Brown} 
\begin{eqnarray}
\left(M_{\mu\rho}^{I}+m^{2}\delta_{\mu\rho}\right)g_{m,\rho\nu}^{I}=\delta_{\mu\nu}-P_{\mu\nu}^{I},\,\,\,\left(M_{\mu\rho}^{I}+m^{2}\delta_{\mu\rho}\right)G_{m,\rho\nu}^{I}=\delta_{\mu\nu},\label{GI}\\
\lim_{m\rightarrow0}g_{m,\mu\nu}^{I}=S_{\mu\nu}^{I}
\end{eqnarray}
where~\cite{Brown} $G_{m,\rho\nu}^{I}=g_{m,\rho\nu}^{I}+\frac{1}{m^{2}}P_{\rho\nu}^{I}.$
It is clear that ${G^{I}}_{m,\mu\rho}^{-1}=\left(M_{\mu\rho}^{I}+m^{2}\delta_{\mu\rho}\right).$

\textbf{Dynamical gluon mass in ILM \cite{ME2017}.} 
Let's repeat the way to Pobylitsa Equation for ILM \char`\"{}scalar\char`\"{}
gluon propagator $\bar{\Delta}$ and neglect by second and higher
orders terms in density expansion. We get 
\begin{eqnarray}
\bar{G}_{m,\rho\nu}-G_{m,\rho\nu}^{0}=N\left(\bar{G}_{m,\alpha\nu}^{I}-{G^{0}}_{m,\rho\nu}\right).\label{G}
\end{eqnarray}
Finally we get at $m\rightarrow0$ limit the dynamical gluon mass
\begin{eqnarray}
M_{g}^{2}\delta_{\rho\nu}=N{S^{0}}_{\rho\sigma}^{-1}(\bar{S}_{\sigma\mu}^{I}-{S^{0}}_{\sigma\mu}){S^{0}}_{\mu\nu}^{-1}\label{Mg}
\end{eqnarray}
From the well-known result for the $S_{\sigma\mu}^{I}$ \cite{BrownCarlitz}
we conclude 
\begin{eqnarray}
M_{g}^{2}\left(q\right)=2M_{s}^{2}\left(q\right).\label{Mg1}
\end{eqnarray}
For the set of parameters~(\ref{eq:SetInstanton}) we estimate that
$M_{g}(0)=362\,MeV.$

\section{Heavy-light quarks interactions in ILM}

\label{heavylight}

If heavy and light quarks are interacting with the same instanton,
they are effectively interacting with each other. Our aim here is
to derive light-heavy quarks interaction term induced by this mechanism.
We solve this taking into account light quarks determinant in the
measure as $D\zeta\Rightarrow D\zeta\,\Det_{low}(\hat{p}+g\hat{A}+im)$.
With account of light quarks ILM heavy quark propagator becomes 
\begin{eqnarray}
\int\prod_{f}D\psi_{f}D\psi_{f}^{\dagger}\exp\int\left(\psi_{f}^{\dagger}(\hat{p}\,+\,im_{f})\psi_{f}\right)\prod_{\pm}\left({\overline{V_{\pm}[\psi^{\dagger},\psi]}}\right)^{N_{\pm}}<T|w[\psi,\psi^{\dagger}]|0>,\label{Qq}
\end{eqnarray}
where 
\begin{eqnarray}
w[\psi,\psi^{\dagger}]=\prod_{\pm}\left({\overline{V_{\pm}[\psi^{\dagger},\psi]}}\right)^{-N_{\pm}}\int D\zeta\,(\theta^{-1}-iA_{4})^{-1}\,\prod_{f}\prod_{\pm}^{N_{\pm}}V_{\pm,f}[\psi^{\dagger},\psi]\label{wpsi}
\end{eqnarray}

The solution of the extended Pobilitca Eq. is \cite{M2014} 
\begin{eqnarray}
 &  & w^{-1}[\psi,\psi^{\dagger}]=\theta^{-1}-\frac{N}{2}\sum_{\pm}\frac{1}{\overline{V_{\pm}[\psi^{\dagger},\psi]}}\Delta_{H,\pm}[\psi^{\dagger},\psi]+O(N^{2}/V^{2}),\\
 &  & \Delta_{H,\pm}[\psi^{\dagger},\psi]=\int d\zeta_{\pm}\prod_{f}V_{\pm,f}[\psi^{\dagger},\psi]\theta^{-1}(w_{\pm}-\theta)\theta^{-1}.\label{DeltaH}
\end{eqnarray}
and it defines the heavy ($Q$)-light quarks($\psi$) interaction
term 
\begin{eqnarray}
S_{Q\psi}=-\lambda\sum_{\pm}Q^{\dagger}\Delta_{H,\pm}[\psi^{\dagger},\psi]Q,\label{SQ}
\end{eqnarray}
where the coupling $\lambda$ is taken from the saddle-point approximation
in light quarks partition function $Z[\xi,\xi^{\dagger}]$ \cite{Musakhanov,Goeke:2007bj}.

\textbf{Heavy\textendash light quarks interactions ($N_{f}=1$) \cite{M2014}.}
In this case the heavy-light quarks interaction term is given by 
\begin{eqnarray}
 &  & S_{Q\psi}=i\int\frac{d^{4}k_{1}}{(2\pi)^{4}}\frac{d^{4}k_{2}}{(2\pi)^{4}}\frac{d^{3}q}{(2\pi)^{3}}(2\pi)^{4}\delta^{3}(\vec{k}_{2}+\vec{k}_{1}-\vec{q})\delta(k_{2,4}-k_{1,4})(M(k_{1})M(k_{2}))^{1/2}\Delta m_{Q}R^{4}\\
 &  & \frac{i_{0}(q\rho)}{i_{0}(0)}\left[\frac{2N_{c}^{2}-N_{c}}{2N_{c}^{2}-2}\psi^{+}(k_{1})\psi(k_{2})Q^{+}Q+\frac{N_{c}^{2}-2N_{c}}{2N_{c}^{2}-2}(\psi^{+}(k_{1})QQ^{+}\psi(k_{2})+\psi^{+}(k_{1})\gamma_{5}QQ^{+}\gamma_{5}\psi(k_{2}))\right]\nonumber 
\end{eqnarray}
The first term is the heavy quark\textendash light meson interaction
term, while the second and the third terms correspond to $Qq$ mesons
degenerated on parity. Similar expression was derived in \cite{Chernyshev:1995gj}.

\textbf{Light quarks contribution to the $Q\bar{Q}$ potential ($N_{f}=1$)
\cite{M2014}. } 
With account of the light quarks averaged previously defined Wilson
loop is given by 
\begin{eqnarray}
 &  & \int\prod_{f}D\psi_{f}D\psi_{f}^{\dagger}\exp\int\left(\psi_{f}^{\dagger}(\hat{p}+im_{f})\psi_{f}\right)\prod_{\pm}\left({\overline{V_{\pm}[\psi^{\dagger},\psi]}}\right)^{N_{\pm}}\tr<T|W[\psi,\psi^{\dagger}]|0>,\label{W1}\\
 &  & <T|W[\psi,\psi^{\dagger}]|0>=\prod_{\pm}\left({\overline{V_{\pm}[\psi^{\dagger},\psi]}}\right)^{-N_{\pm}}\int D\zeta\prod_{f}\prod_{\pm}^{N_{\pm}}V_{\pm,f}[\psi^{\dagger},\psi]\nonumber \\
 &  & \times P\exp(i\int_{L_{1}}dx_{4}A_{4})P\exp(i\int_{L_{2}}dx_{4}A_{4})\label{W2}
\end{eqnarray}

In the first order on instanton density the solution of extended Pobilitca
Eq. is given by 
\begin{eqnarray}
 &  & W^{-1}[\psi,\psi^{\dagger}]=w_{1}^{-1}[\psi,\psi^{\dagger}](\times)w_{2}^{-1,T}[\psi,\psi^{\dagger}]-\frac{N}{2}\sum_{\pm}\left({\overline{V_{\pm}[\psi^{\dagger},\psi]}}\right)^{-1}\nonumber \\
 &  & \times\int d\zeta_{\pm}\prod_{f}V_{\pm,f}[\psi_{f}^{\dagger},\psi_{f}]\left(\theta^{-1}\left(w_{\pm}^{(1)}-\theta\right)\theta^{-1}\right)(\times)\left(\theta^{-1}\left(w_{\pm}^{(2)}-\theta\right)\theta^{-1}\right)^{T}+O(\frac{N^{2}}{V^{2}})\label{W3}
\end{eqnarray}
where the superscript $T$ means the transposition and $(\times)$
stands for the tensor product.

\begin{figure}[h]
\centering \includegraphics[scale=0.2]{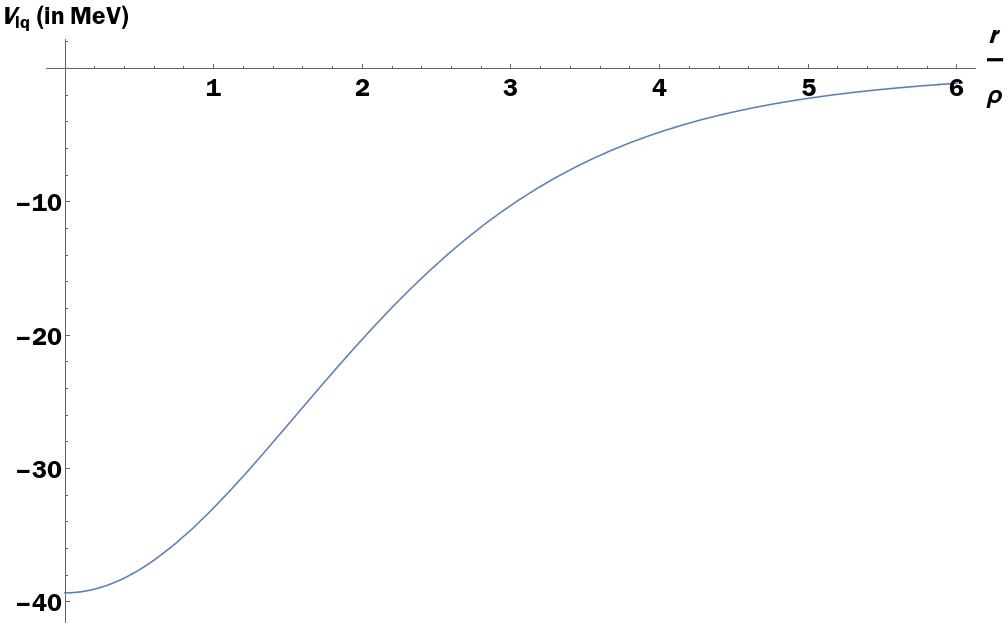}\caption{ Heavy quark\textendash antiquark potential $V_{lq}(r/\rho)$ (in
MeV), generated by light quarks, at Set~II $\rho=0.36$~fm, $R=0.89\,\mathrm{fm}$.}
\label{Vlq} 
\end{figure}

It is obvious that first term in \re{W3} describes light quarks
exchange between heavy quarks, which leads to the potential $V_{lq}(r/\rho)$
\cite{M2014}, see Fig.\ref{Vlq}. 

\textbf{$N_{f}=2$. Heavy quark light mesons interaction term \cite{M2014}.}
Heavy quark light quarks interaction term~\re{SQ} has an essential
part which is co-product of colorless heavy and light quark factors.
From this one and at the saddle points we have the effective action
for the mesons and colorless heavy quark $Q^{\dagger}Q$ bilinear
as 
\begin{eqnarray}
 &  & S[\sigma',\vec{\phi}',\eta',\vec{\sigma}',Q^{\dagger}Q]=-\Tr\ln\frac{\hat{p}+i(m+M(p))}{\hat{p}+im}+N/2+\frac{1}{2}\int d^{4}x\left({\sigma'}^{2}+{\vec{\phi'}}^{2}+{\vec{\sigma'}}^{2}+{\eta'}^{2}\right)\nonumber \\
 &  & -\Tr\ln\left[1+\frac{1}{\hat{p}+i(m+M(p))}\frac{iM}{\sigma_{0}}F\left(\sigma'+i\gamma_{5}\vec{\tau}\vec{\phi}'+i\vec{\tau}\vec{\sigma}'+\gamma_{5}\eta'\right)F\right]\label{SQmesons}\\
 &  & +\Tr\frac{1}{\hat{p}+i(m+M(p))+\frac{iM}{\sigma_{0}}F\left(\sigma'+i\gamma_{5}\vec{\tau}\vec{\phi}'+i\vec{\tau}\vec{\sigma}'+\gamma_{5}\eta'\right)F}\nonumber \\
 &  & \times i\left(M(p)+\frac{M}{\sigma_{0}}F\left(\sigma'+i\gamma_{5}\vec{\tau}\vec{\phi}'+i\vec{\tau}\vec{\sigma}'+\gamma_{5}\eta'\right)F\right)\left(\frac{i}{2}\Delta m_{Q}R^{4}\int e^{-ipx}\frac{d^{4}p_{1}}{(2\pi)^{4}}\frac{d^{4}p_{2}}{(2\pi)^{4}}\frac{i_{0}(p\rho)}{i_{0}(0)}Q^{\dagger}Q\right).\nonumber 
\end{eqnarray}
The first and the second lines describe  mesons and their interactions,
while the third and the forth one describe the renormalization of
the heavy quark mass and heavy quark-light quark mesons interactions
terms. From the Eq.~(\ref{SQmesons}) we have the heavy quark-pion
interaction term:

\begin{eqnarray}
S_{Q\pi}=i\Delta m_{Q}R^{4}\frac{F_{\pi Q}^{2}}{4}\int d^{4}x\,\tr_{f}\partial_{\mu}U(x)\partial_{\mu}U^{\dagger}(x)\int e^{-ipx}\frac{d^{4}p_{1}}{(2\pi)^{4}}\frac{d^{4}p_{2}}{(2\pi)^{4}}\frac{i_{0}(p\rho)}{i_{0}(0)}Q^{\dagger}(p_{2})Q(p_{1})\label{SQpions3}
\end{eqnarray}
where the matrix pion field $U\approx(\sigma+i\vec{\tau}\vec{\phi}')/\sigma_{0}$,
$F_{\pi Q}^{2}\approx0.7F_{\pi}^{2}$ and $p=p_{1}-p_{2}.$

The similar approach for the calculations of $Q\bar{Q}$ correlators
(Wilson loop) with account of light quarks will provide the interaction
term of the pair of heavy quarks with pions $S_{QQ\pi}$. Both of
these terms $S_{QQ\pi}$ and $S_{Q\pi}$ are responsible for the two-pions
transitions in heavy quarkoniums. 

\section{Discussion and future work}

\label{discussion}
\begin{itemize}
\item Instantons with sizes $\rho$ $\sim$ hadron quark core sizes $r$
give most essential contribution to their properties. In the case
of lowest on energy hadron states the ILM is applicable.
\item The strength of a heavy quark-instanton interaction is defined by

$\Delta m_{Q}\sim packing\,\,fraction\,\,\rho^{-1}\sim70\,MeV$ (at
$\rho=0.33\,fm,R=1\,fm$)

is small, while 
 the strength of a light quark-instanton and gluon-instanton interactions
are much more larger and given by the dynamical quark and dynamical
gluon masses

$M_{g}\sim M\sim(packing\,\,fraction)^{1/2}\,\,\rho^{-1}\approx360\,MeV$
(at the same $\rho,R$).



\item Instantons naturally generate also heavy-light quarks interaction,
which might be important for the heavy quarkonium and heavy-light
quarks systems properties. It can be responsible for the traces of
SBCS in heavy quarks physics.
\end{itemize}
\textbf{Future work. } 
\begin{itemize}
\item Extend the calculations of heavy-heavy quarks potential with ILM modified
gluons. 
\item Take into account light quarks in the observables of heavy quark physics: 
\begin{itemize}
\item $Q\bar{Q}$ pions transitions; 
\item heavy-light mesons etc. 
\end{itemize}
\item Consider ILM generated gluon-light quarks interactions and related
problems of exotic hadrons. 
\end{itemize}



\textbf{Acknowledgement.} M.M.  {is} thankful to Marat Siddikov for the useful
 and helpful communications. This work is supported by Uz grant
 OT-F2-10. 



\end{document}